\def\arcsec{$^{\prime\prime}$}
\def\arcsecl{$^{\prime\prime}$ }

\documentclass[]{spie}  
\usepackage[]{graphicx}

\title{The Development of WIFIS: a Wide Integral Field Infrared Spectrograph } 

\author{Suresh Sivanandam\supit{a}, Richard C. Y. Chou\supit{b}, Dae-Sik Moon\supit{b}, Ke Ma\supit{b}, Maxwell Millar-Blanchaer\supit{b}, Stephen S. Eikenberry\supit{c}, Moo-Young Chun\supit{d}, Sang Chul Kim\supit{d}, Steven N. Raines\supit{c}, Joshua Eisner\supit{e}
\skiplinehalf
\supit{a}Dunlap Institute, University of Toronto, 50 St. George St, Toronto, ON, Canada; \\
\supit{b}Department of Astronomy, University of Toronto, 50 St. George St, Toronto, ON, Canada; \\
\supit{c}Department of Astronomy, University of Florida, 211 Bryant Space Science Center, Gainesville, FL, USA; \\
\supit{d}Korea Astronomy and Space Science Institute, 776 Daedeokdae-ro, Yuseong-gu, Taejon, Korea; \\
\supit{e}Steward Observatory, University of Arizona, 933 N. Cherry Ave, Tucson, AZ, USA
}

\authorinfo{Further author information: (Send correspondence to S.S.)\\S.S.: E-mail: sivanandam@di.utoronto.ca, Telephone: +1 (416) 978-2183}

\begin{document} 
  \maketitle 

\begin{abstract}

We present the current results from the development of a wide integral field infrared spectrograph (WIFIS). WIFIS offers an unprecedented combination of etendue and spectral resolving power for seeing-limited, integral field observations in the $0.9-1.8$ $\mu$m range and is most sensitive in the $0.9-1.35$ $\mu$m range. Its optical design consists of front-end re-imaging optics, an all-reflective image slicer-type, integral field unit (IFU) called FISICA, and a long-slit grating spectrograph back-end that is coupled with a HAWAII 2RG focal plane array. The full wavelength range is achieved by selecting between two different gratings. By virtue of its re-imaging optics, the spectrograph is quite versatile and can be used at multiple telescopes. The size of its field-of-view is unrivalled by other similar spectrographs, offering a 4.5\arcsec$\times$ 12\arcsecl integral field at a 10-meter class telescope (or 20\arcsec$\times$ 50\arcsecl at a 2.3-meter telescope). The use of WIFIS will be crucial in astronomical problems which require wide-field, two-dimensional spectroscopy such as the study of merging galaxies at moderate redshift and nearby star/planet-forming regions and supernova remnants. We discuss the final optical design of WIFIS, and its predicted on-sky performance on two reference telescope platforms: the 2.3-m Steward Bok telescope and the 10.4-m Gran Telescopio Canarias. We also present the results from our laboratory characterization of FISICA. IFU properties such as magnification, field-mapping, and slit width along the entire slit length were measured by our tests. The construction and testing of WIFIS is expected to be completed by early 2013. We plan to commission the instrument at the 2.3-m Steward Bok telescope at Kitt Peak, USA in Spring 2013.

\end{abstract}



\section{INTRODUCTION}
\label{sec:intro}  
\begin{figure}
   \begin{center}
   \begin{tabular}{c}
   \includegraphics[height=7cm]{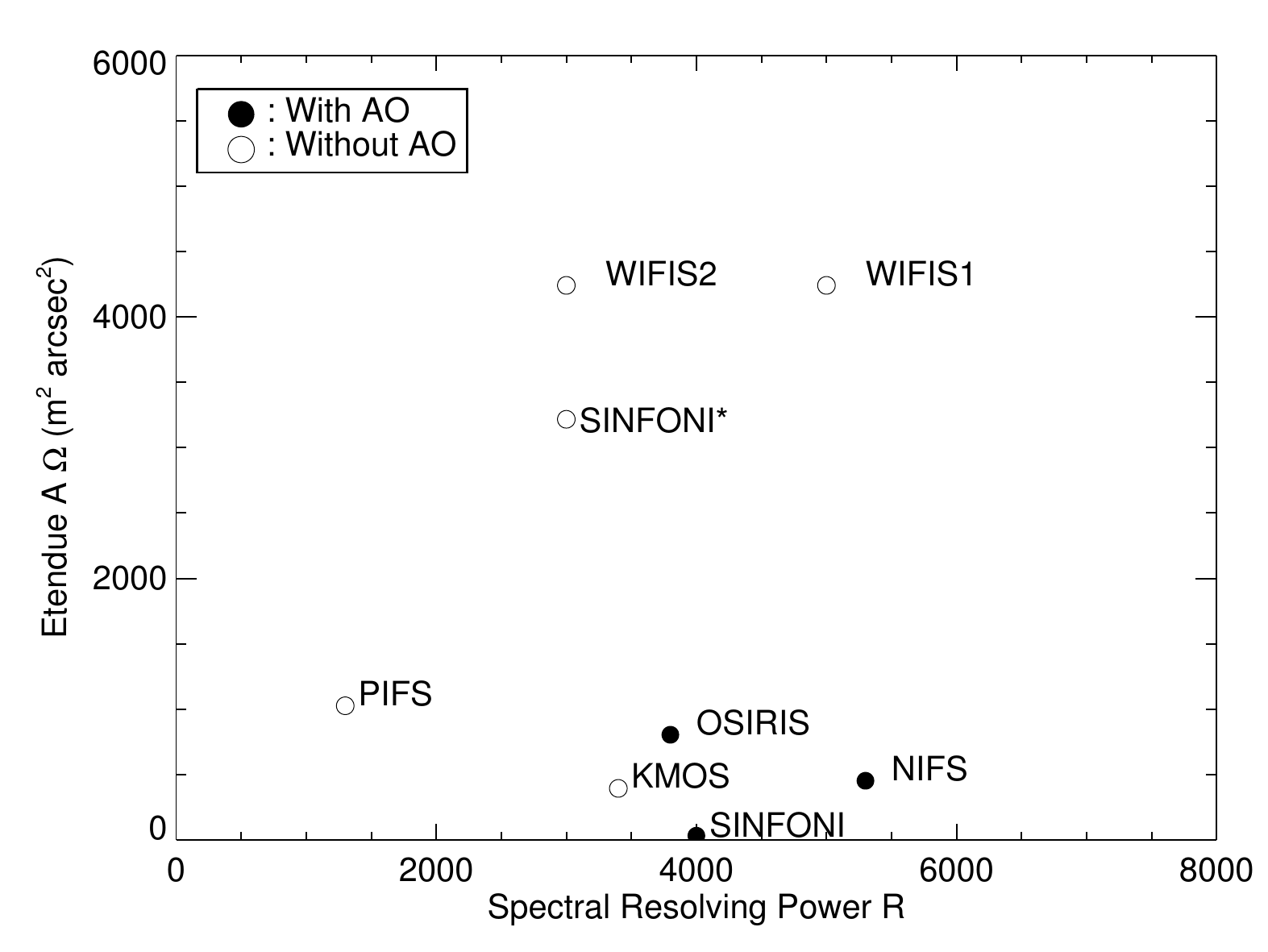}
   \end{tabular}
   \end{center}
   \caption[ifuetendue] 
   { \label{fig:ifuetendue} 
   The etendue values of several near-IR integral field spectrographs currently in use or ones that will be commissioned in the near future such as KMOS. We also show the expected etendue of our previous, WIFIS1, and current, WIFIS2, designs. SINFONI can be used in both AO and non-AO mode, where its non-AO mode etendue is shown with an asterix. KMOS has 24 IFUs that can be deployed across a large field. We only show the etendue of a single KMOS IFU because there are mechanical restrictions on how close the individual IFUs can be deployed. WIFIS clearly has an unprecedented combination of etendue and spectral resolving power over a single field. 
}
   \end{figure}

With the advent of megapixel, CMOS-based detectors, integral field spectroscopy (IFS) has become practicable in the near-infrared (NIR). IFS allows one to obtain spectra over both spatial dimensions, providing spectral information at each spatial co-ordinate on the sky, which is crucial for the study of various types of interesting astronomical objects (e.g., kinematics, chemistry, and stellar properties of high-redshift galaxies). However, most NIR integral field spectrographs used in large telescopes are coupled with adaptive optics (AO) systems to produce high spatial resolution, integral field spectra at moderate spectral resolution ($R \sim 3000$). As a result, the etendue ($A\times\Omega$), a figure of merit that reflects how efficiently a telescope is used per unit time, of these spectrographs is relatively small. In Figure \ref{fig:ifuetendue} we compare the etendues and spectral resolving powers of current/upcoming NIR integral field spectrographs such as KMOS, along with the values for our own Wide Integral Field Infrared Spectrograph (WIFIS) instrument that we describe in this paper. Most of these spectrographs predominantly focus on scientific problems that require high angular resolution and small fields at moderate spectral resolution. However, there are a myriad of scientific problems that require IFS over a larger field. These include the characterization of the ionization properties of gas in nearby extended star forming regions or supernova remnants, the determination of the kinematics and stellar properties of nearby galaxies, and the measurement of kinematics and star-forming properties of moderate redshift ($z \sim 0.5-1.0$) merging galaxies. In all of these cases, the etendue of the instrument becomes critical because one needs to obtain spectroscopic information of fairly extended and faint emission. For this reason, we set out to construct WIFIS, an integral field spectrograph with an unprecedented etendue (see Figure \ref{fig:ifuetendue}).

\par
WIFIS is an image slicer-based imaging spectrograph designed to have the maximum size of the integral field affordable with the Hawaii 2RG (H2RG) NIR array of $2048\times2048$ pixels with a broad-band spectral coverage at a medium spectral resolution (R $\sim 3,000-5,000$) in a single exposure.
 The desire to maximize both spatial and spectral coverage while maintaining satisfactory optical performance is quite challenging, resulting in a very fast optical system for the spectrograph camera (see below). Two separate optical designs were developed for WIFIS: (1) WIFIS1 was designed to be a cryogenic instrument that operated over the $0.9-2.5\mu$m range with a spectral resolving power R$\sim 5000$ and a field-of-view of 4.5\arcsec$\times$ 12\arcsec at a 10-meter telescope. The full spectral range was not accessible simultaneously and different gratings had to be used to obtain complete spectral coverage. The optical design of this instrument is discussed in our previous work\cite{chou2010}. (2) WIFIS2 does not require fully cryogenic optics and maintains the same field-of-view (and etendue), but has a reduced resolving power of $R\sim3000$ and operating wavelength range. The WIFIS2 design operates over the $0.9-1.8$ $\mu$m spectral range with the $0.9-1.35$ $\mu$m bandpass offering the best spectral resolving power and sensitivity. The full spectral coverage is also achieved using two different gratings. In the interest of reducing the complexity, cost of the spectrograph, and deployment time, we chose to construct our WIFIS2 design. Hereafter, we will refer to the WIFIS2 design as WIFIS. 

   \begin{figure}
   \begin{center}
   \begin{tabular}{c}
   \includegraphics[height=7cm]{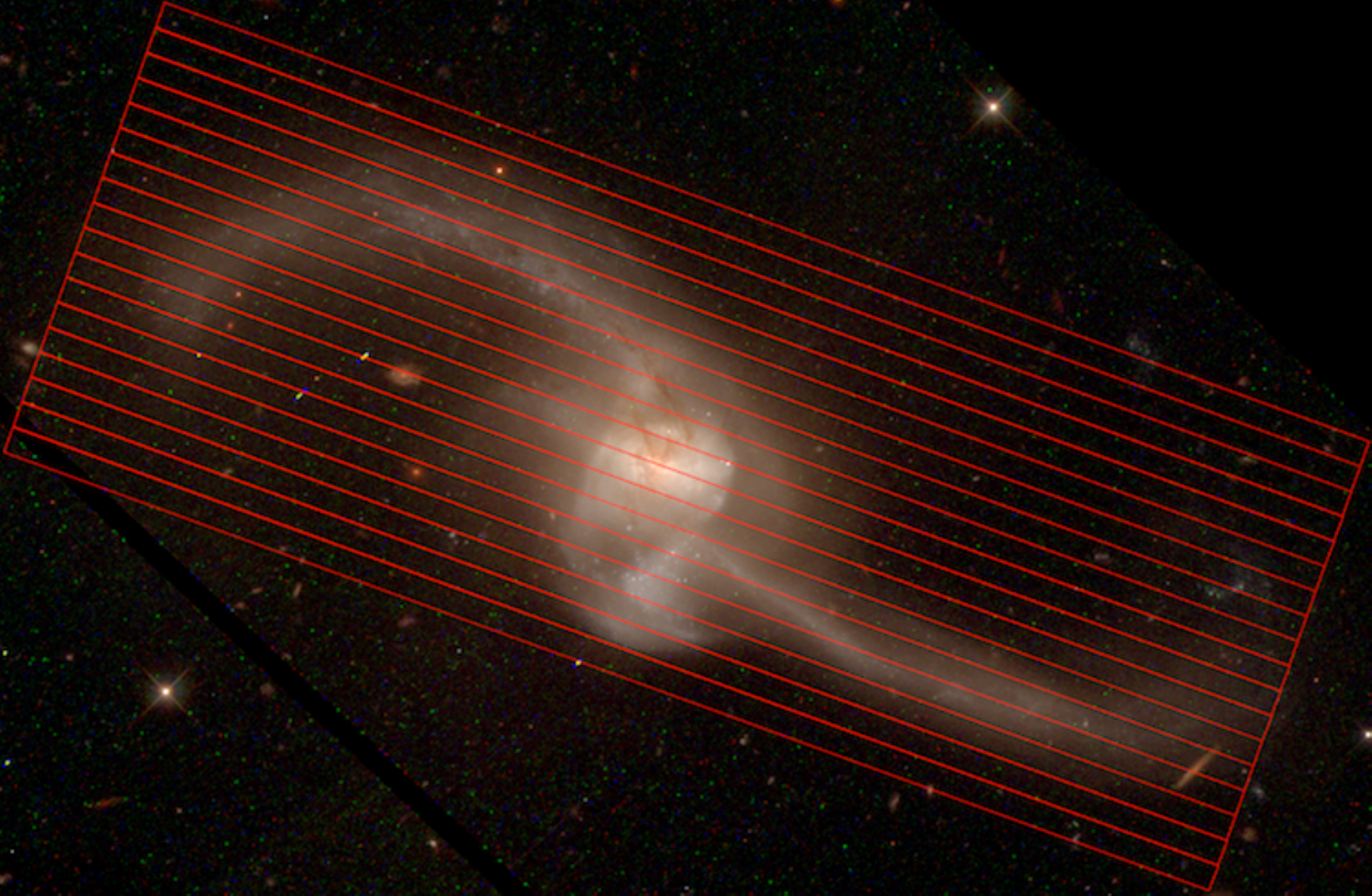}
   \end{tabular}
   \end{center}
   \caption[ngc2623] 
   { \label{fig:ngc2623} 
    An optical Hubble Space Telescope colour composite of NGC 2623, a nearby major merger of galaxies. The intensities are shown in a logarithmic scale to reveal the low surface brightness features. We overlay the expected 4.5\arcsec$\times$12\arcsecl field-of-view of WIFIS if this galaxy was at a redshift of 0.5, the distance at which H$\alpha$ emission is visible in the WIFIS bandpass. Each red rectangular region is a 0.25\arcsec$\times$ 12\arcsecl slice of WIFIS's integral field when coupled with a 10-meter telescope. One can see the benefits of the large etendue afforded by WIFIS when observing merging galaxies at moderate redshift. 
}
   \end{figure}
\par
WIFIS has been designed to be versatile and can operate at different telescopes by simply replacing its reimaging optics. Its immediate intended destination is the Steward Bok 2.3-meter telescope. While the aperture size is relatively small, the size of its integral field-of-view is notably large (20\arcsec$\times$ 50\arcsec) and is perfectly suited for mapping line emission (e.g. Pa$\beta$, [Fe II]) from extended galactic star forming regions, supernova remnants, and measuring the kinematics and stellar properties of nearby galaxies. For a 10-m class telescope, such as the 10.4-meter Gran Telescopio Canarias (GTC), WIFIS has a smaller field (4.5\arcsec$\times$12\arcsec) with the increased sensitivity optimized to study the star forming properties of merging galaxies at moderate redshift $z \sim 0.5-1.0$ where the H$\alpha$ line is redshifted into WIFIS's most sensitive bandpass. An example observation is shown in Figure \ref{fig:ngc2623} where a local major merger, NGC 2623, has been placed at a redshift of $z \sim 0.5.$ The large field-of-view of WIFIS is well-matched to the physical size of this merger, which clearly illustrates the necessity of large field IFS in the NIR wavebands. 
\par
In this paper, we discuss the final optical design and system layout of WIFIS along with its predicted performance, including its on-sky characteristics, sensitivity, and spectral resolving power at both the Bok telescope and the GTC. We also present the results from the laboratorial characterization of its integral field unit. Finally, we discuss the current status of the project and our future plans. 

\section{System Design and Optical Performance} 
\label{sec:opticaldesign}

\subsection{Optical Design}
\par
\begin{figure}
   \begin{center}
   \begin{tabular}{c}
   \includegraphics[height=9cm]{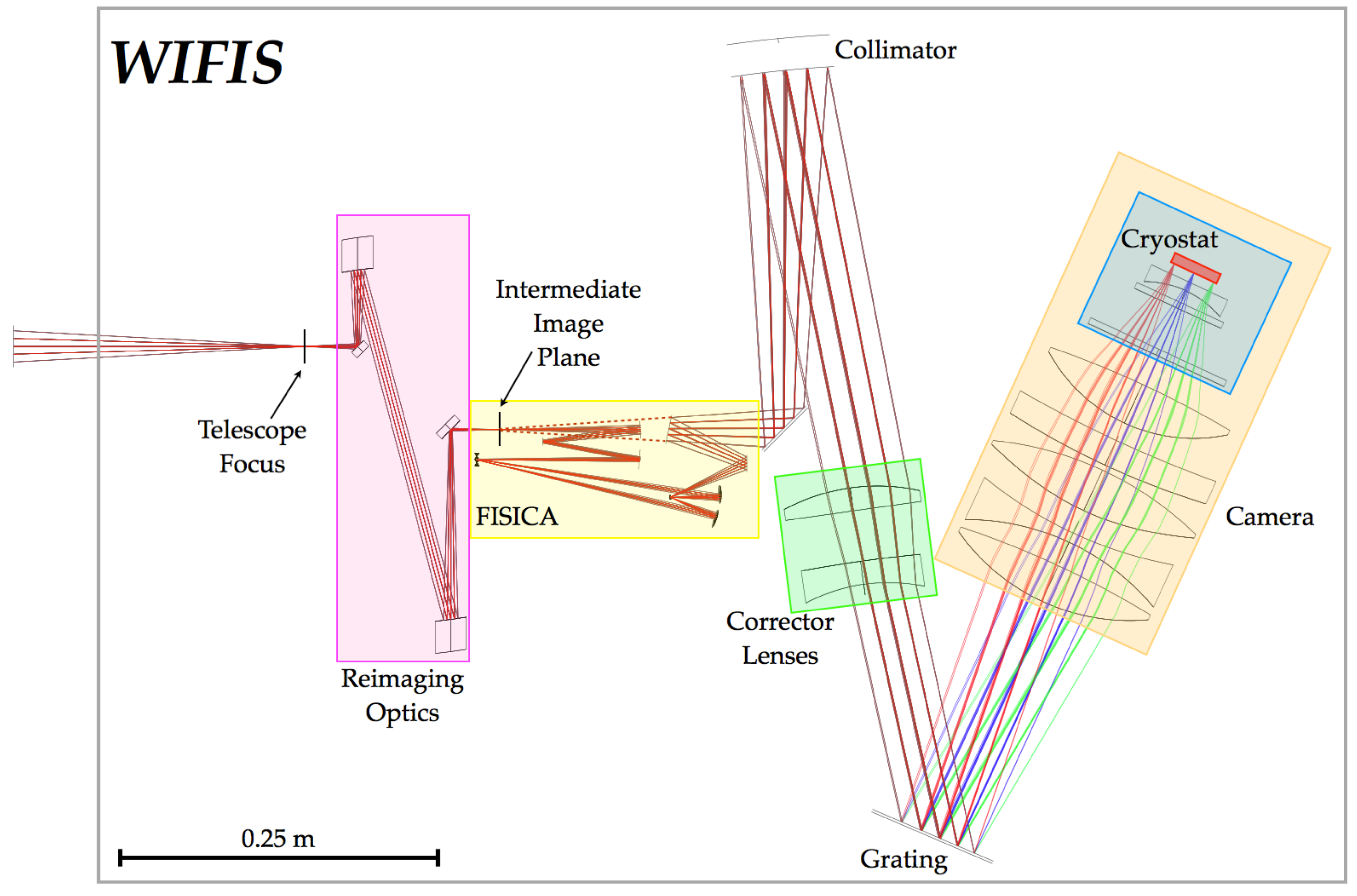}
   \end{tabular}
   \end{center}
   \caption[opticslayout] 
   { \label{fig:opticslayout} 
The optical layout of the revised WIFIS design. The major components of the spectrograph are highlighted. The reimaging optics is used to change the focal ratio of the input beam to match the one required by FISICA, an image slicer IFU. The intermediate image plane of the reimaging optics is shown in the diagram. FISICA reformats the input rectangular field into a long slit, which is in the form of a virtual image also located at the intermediate image plane. The marginal rays emanating from the virtual slit are shown by the dashed lines drawn inside FISICA. The collimation optics consist of a collimating mirror and two correction lenses. One of the two lenses has an aspheric surface. The grating can be swapped between $J$-band or $H$-band gratings depending on the wavelength of operation. The final set of lenses make up the f/3 camera. The front surface of the third lens is aspheric. The detector, the final camera lens of camera, and the thermal blocking filter are housed within a cryostat. The H2RG detector is shown as a red rectangle.}
   \end{figure}

The optical design of WIFIS consists of five major components: (1) reimaging optics, (2) image slicer-based integral field unit (IFU), (3) collimation optics, (4) grating, and (5) spectrograph camera. The optical layout of the spectrograph is shown in Figure \ref{fig:opticslayout} and is very similar to the original WIFIS1 design\cite{chou2010}. The IFU has already been constructed as part of a previous research program\cite{glenn2004}. Considerable effort went into reducing the manufacturing cost of the instrument by minimizing the number of aspheric surfaces and using relatively inexpensive lens materials in the optical design. The final design consists of mostly spherical surfaces with the exception of the reimaging optics and two aspheric lens surfaces. For the reimaging optics, we use a pair of off-axis parabolic mirrors. In order to control aberrations, we introduced an aspheric surface on one of the collimator corrector lenses and another one in the spectrograph camera. All reflective components are gold-coated, while the refractive optics have an anti-reflection coating that has a reflectivity of $< 1.5$\% in the $0.9-1.8\mu$m range. 

\par
The reimaging optics are custom fabricated two off-axis parabolas that convert the telescope input beam to an f/16 beam and create an intermediate image plane at the input of the IFU. The IFU, called Florida Image Slicer for Infrared Cosmology and Astrophysics (FISICA), is an image slicer-type IFU that was constructed by the University of Florida\cite{glenn2004,eikenberry2004,eikenberry2006}. For seeing-limited observations, the image slicer-type IFU is best suited for wide-field operation compared to a lenslet array-based IFU because of its coarse spatial sampling and higher throughput. For AO, however, the image slicer-type IFU is not well-suited because it introduces more wavefront errors compared to lenslet array-type IFU made with low index of refraction material. Another disadvantage of image slicer-based IFU is its packaging because it is much more bulky than a lenslet array-type IFU. In our case, the image slicer-based IFU is best suited for our instrument requirements.
\par
FISICA is a fully reflective design, which can also be used under cryogenic conditions. The IFU's 22-element image slicer slices a rectangular field of $9.1 \times 4.3$ mm in size, located at the intermediate image plane, and converts it into a $100\times 0.098$ mm pseudo long slit.  The slicer also demagnifies the beam and produces an f/8 output beam. The optical layout of FISICA is fairly complex. The telescope field is reimaged onto the slicer array using a three mirror relay system. After the field is sliced, the light passes through a pupil mirror array, a field mirror array, and two different fold mirrors, which reformat the light from the 22 slices into a single pseudo long slit. The pseudo long slit is a virtual image at the same location as the intermediate image plane, so the optics downstream of the IFU share the same design as a long slit spectrograph. Only 18 of the 22 slices are used in the WIFIS design.
\par
\begin{figure}
   \begin{center}
   \begin{tabular}{c}
   \includegraphics[height=8cm]{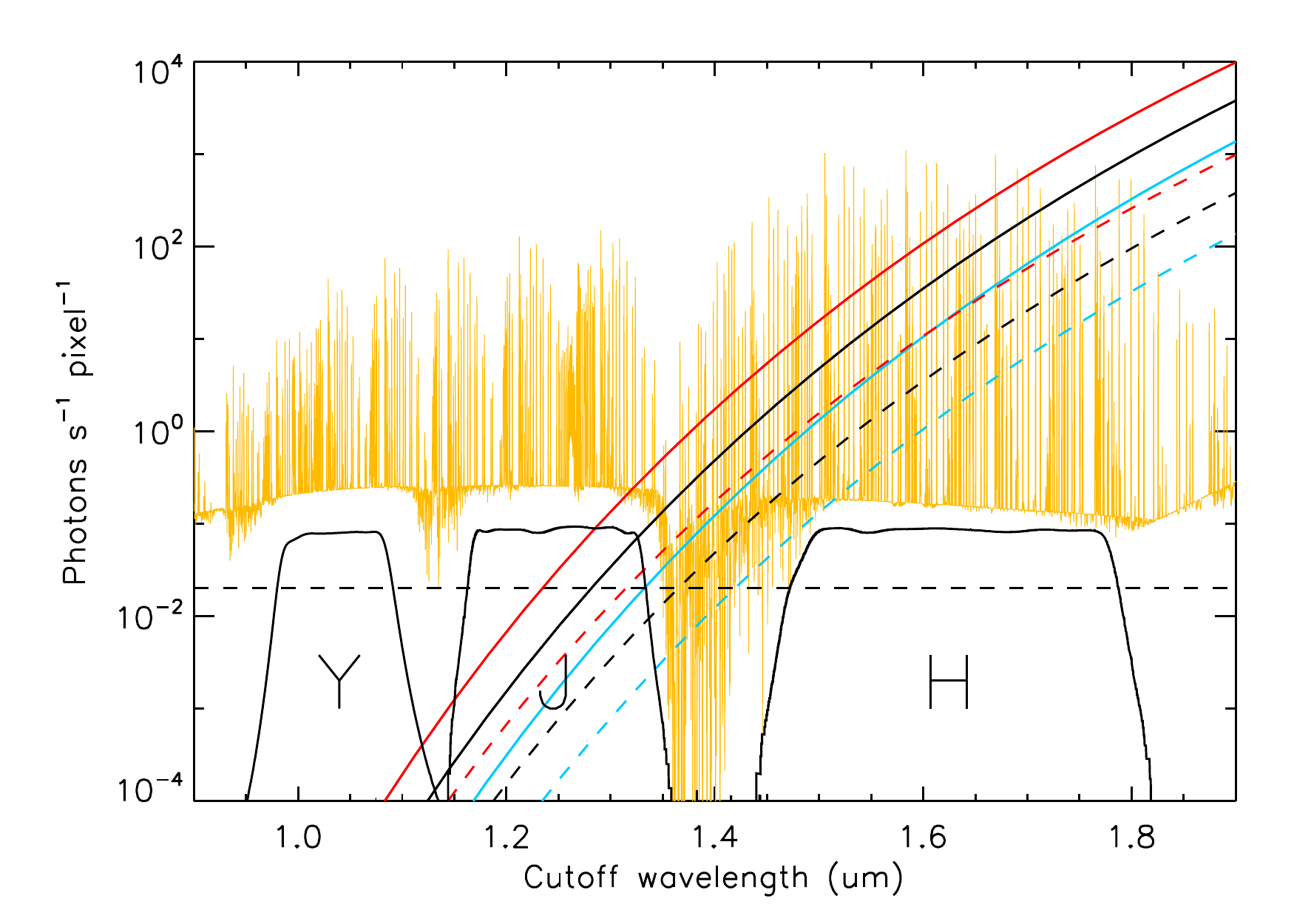}
   \end{tabular}
   \end{center}
   \caption[thermal]
   { \label{fig:thermal} 
   Expected photon flux per pixel from thermal radiation as a function of the cut-off wavelength of the thermal blocking filter used within instrument. The solid and dashed curves are thermal fluxes for a blackbody and a grey body with an emissivity of 0.1, respectively. The blue, black, and red curves represent thermal emission from 0, 10, 20$^\circ$ Celcius bodies, respectively. The orange curve is model of the dispersed sky emission convolved to the spectral resolution of WIFIS. The common near-IR filter bandpasses are shown for reference. The dashed horizontal line is the approximate dark current for an H2RG in photon units. Thermal light begins to dominate long ward of $J$-band and is the main source of background in $H$-band. 
   }
   \end{figure}

The collimator system consists of a large spherical mirror and two correction lenses. One of the correction lens has a single aspherical surface to control aberrations. The gratings are off-the-shelf components from Newport and are mounted on a translation stage. We will be using one grating optimized for  $0.9-1.35$ $\mu$m and another for $H$-band operation. Both gratings will operate in $m=1$ mode to maximize efficiency. Finally, the most complicated component of this instrument is the f/3 spectrograph camera. The optical train of the spectrograph camera consists of six lenses, a sapphire window and a filter. The last lens, the filter, and the H2RG detector are located inside a cryostat that is cooled by liquid nitrogen and kept at 77K. The filter is housed within a translation stage that allows one to switch among a closed position for taking darks, a thermal blocking filter that blocks out light longer than $1.35\mu$m, and an $H$-band filter. Since the H2RG is sensitive out to $\sim 2.5\mu$m, a blocking filter is required to ensure thermal light produced by the instrument, telescope, and sky is blocked. Our thermal emission analysis of the instrument, shown in Figure \ref{fig:thermal}, suggests that WIFIS will be thermal photon noise limited in the $H$ band even in the most optimistic case where we see grey body (constant emissivity as a function of wavelength) emission at relatively low ambient temperatures. We discuss this further in Section \ref{subsec:thermal}.
 
\subsection{Optical Performance and Spectral Resolving Power}
\par
\begin{figure}
   \begin{center}
   \begin{tabular}{c}
   \includegraphics[height=6cm]{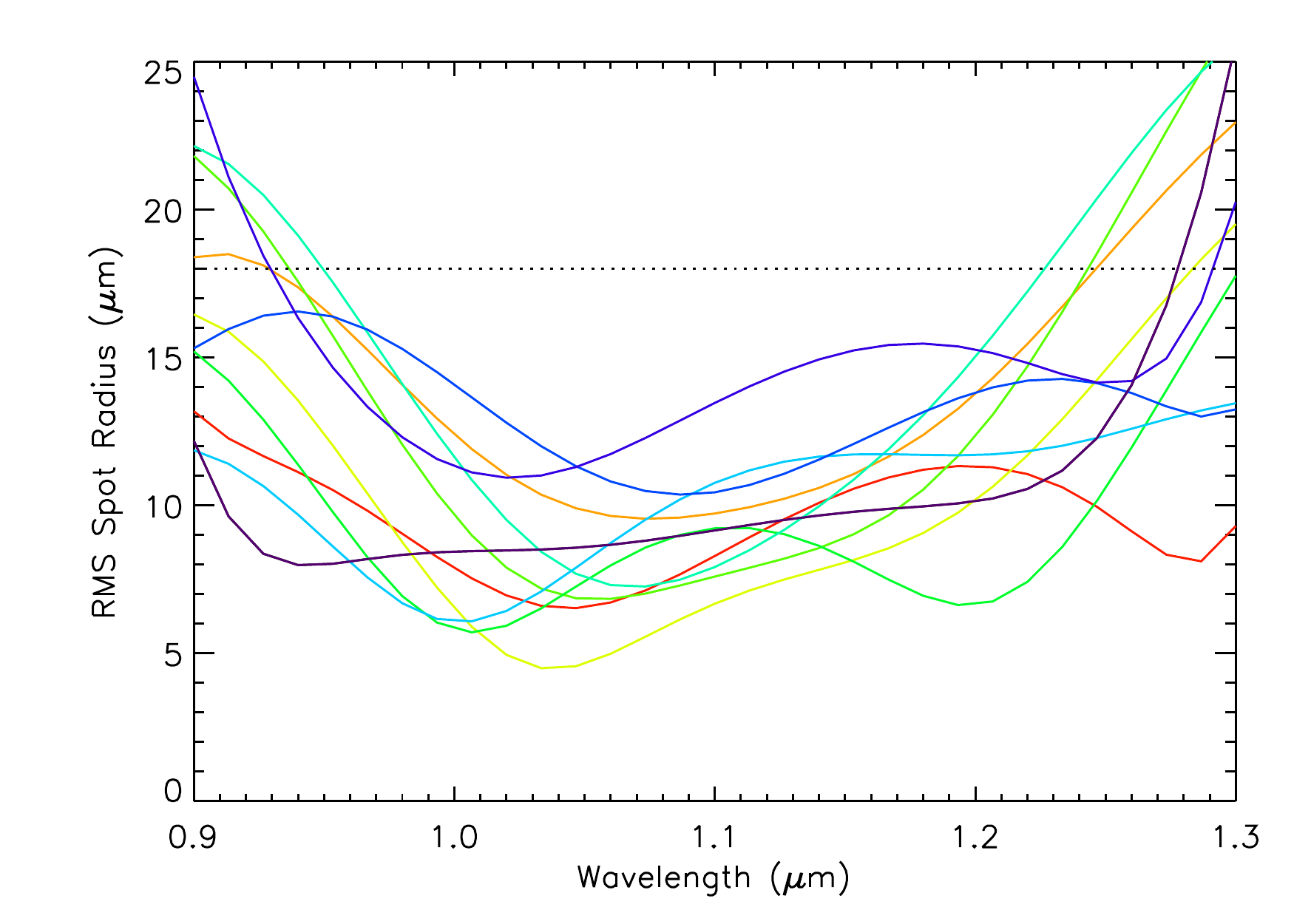} \includegraphics[height=6cm]{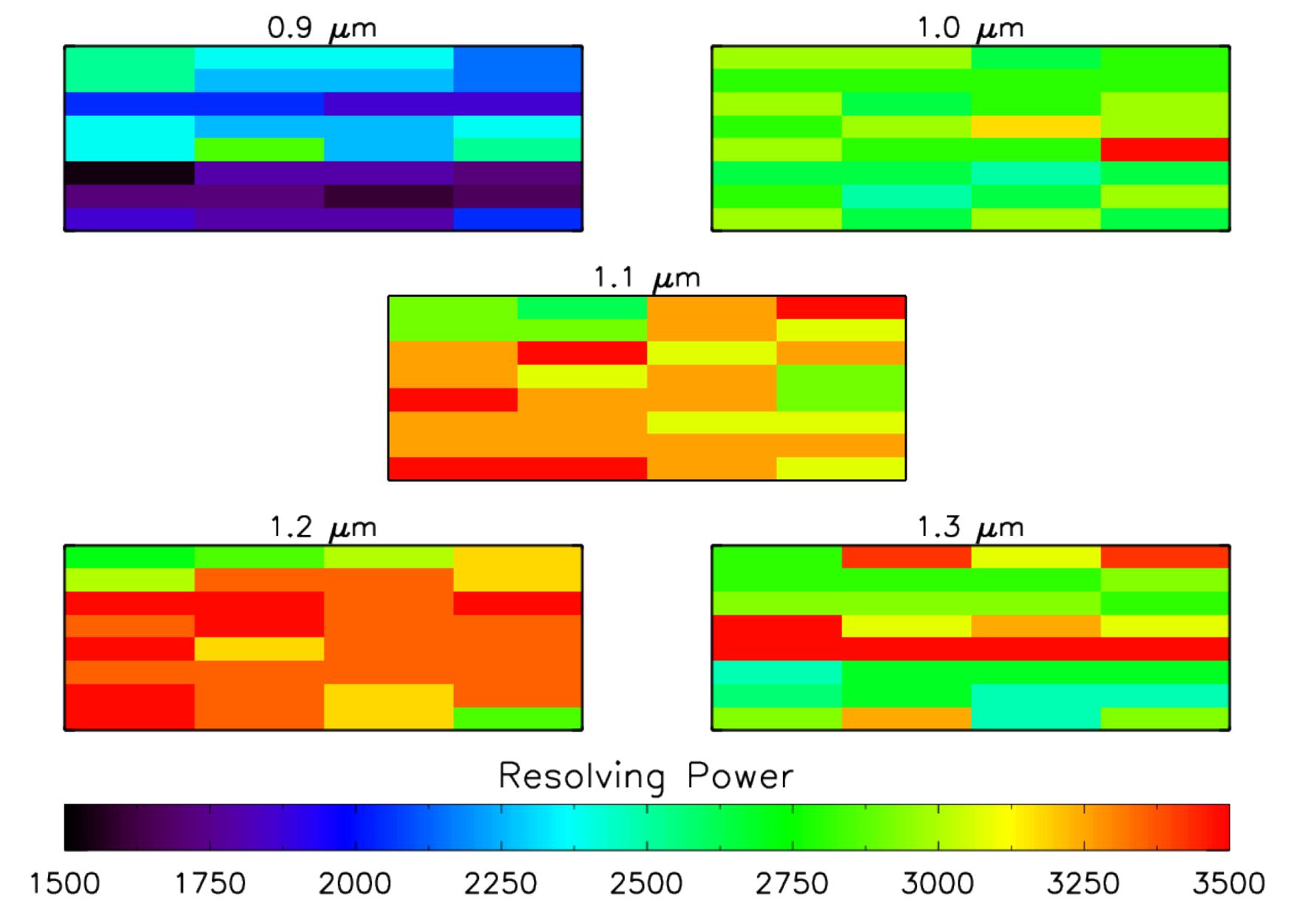}
   \end{tabular}
   \end{center}
   \caption[spotdiagram]
   { \label{fig:spotdiagram} 
   {\it Left:} Imaging performance in terms of RMS spot radius for multiple positions across the IFU field as a function of wavelength. The width of H2RG pixel is shown by the dotted line. Our requirement was to achieve a spot radius of at most one pixel across the full wavelength ($0.9-1.3$ $\mu$m) range and we largely meet this requirement for all of our fields. {\it Right:} Simulated spectral resolving power as a function of IFU field position for seeing-limited observations. Each rectangle represents the spectral resolving power for the entire rectangular IFU field for a given wavelength. In general, for wavelengths longer than 1.0 $\mu$m, we reach our goal of $R\sim3,000$ across the entire field.
}
   \end{figure}

We have modelled the optical performance of WIFIS to confirm that it meets our spectral resolving power and image quality requirements. The driving requirement is the spectral resolving power, which requires the long slit image on the spectrograph camera to span around 2 pixels on the detector. As shown in the left panel of Figure \ref{fig:spotdiagram}, the RMS spot size as predicted by ZEMAX for multiple field positions generally falls within our requirement over the $0.9-1.3\mu$m wavelength range. We also simulated the real system performance by feeding WIFIS coupled to a telescope with a Gaussian point spread function with a full-width-half-max (FWHM) $ = 1.0$\arcsec. The system performance can be obtained by measuring the FWHM of the slit image in the spectral direction on the detector plane at various wavelengths through the \emph{image-simulation} function provided by ZEMAX. The slit FWHM defines the spectral coverage of a resolution element $d\lambda$ and the resolving power $R$ is then calculated by $\lambda/ d\lambda.$ The results of this simulation are shown in the right panel of Figure \ref{fig:spotdiagram} where the spectrograph has a resolving power of $\sim3,000$ over the $0.9-1.3\mu$m range. In the H-band, however, its resolving power degrades significantly to $R\sim1,500$ due to residual chromatic aberrations.
\subsection{Thermal Effects}
\label{subsec:thermal}
\par
From our modelling, it became clear that thermal effects need to be carefully accounted for. There are two separate effects that need to be considered: (1) The variation in optical performance due to temperature changes in the optical train; (2) Thermal light from the uncooled optics and the instrument enclosure that is seen by the detector. Both of these issues need to be carefully addressed to ensure that we obtain the necessary optical performance and overall sensitivity.  
\par
To test if the system performance is significantly degraded by the expected seasonal/diurnal fluctuations in ambient temperature, we carried out a ZEMAX simulation of the spot size variation by adjusting the global system temperature.  For typical seasonal/diurnal variations in temperature within the range of 0 to 30$^\circ$ Celcius (C), a significant change in optical performance was observed, with approximately a factor of two change in RMS spot size for a 10$^\circ$C variation in temperature. Our solution for this issue is to simply shift the detector focus by translating the cryostat along the optical axis. A translation of approximately 150 $\mu$m, with an accuracy of $50\mu$m was sufficient to remove this effect. The shifting of the cryostat also had a negligible effect on the wavelength mapping of the spectrograph. In our final design, we include a translation stage for the cryostat that actively shifts its position to compensate for temperature fluctuations throughout the night. Of course, this only accounts for a global change in temperature. It is possible to have temperature gradients across the optical train. To mitigate this effect, an insulating enclosure for the instrument using insulating foam will constructed and all potential heat sources within the uncooled part of the instrument will be removed. 
\par
A more insidious problem is the thermal light generated by the uncooled enclosure and optics. The cut-off wavelength of the detector is approximately $2.5\mu$m. This means that a significant amount of thermal light produced by the enclosure and optics outside the cryostat will be seen by the detector. The unfortunate aspect of this issue is that the thermal light component will not be dispersed like the sky background because a majority of it will come from the spectrograph camera and enclosure. As mentioned earlier, to combat this issue we will be using a custom-made thermal blocking filter to block out light redder than our operating wavelength range. The degree to which thermal light affects our observations was simulated. The results of this simulation are shown in Figure \ref{fig:thermal}. The sky emission (orange curve) has been appropriately convolved to match our spectrograph resolution and multiplied by the end-to-end throughput of the telescope and the instrument. The solid and dashed curves show the photon flux observed by a detector pixel if there was a cut-off filter upstream that removed all light longward of the specified wavelength. We test a range of instrument emissivities and temperatures: the dashed curves represent an optimistic emissivity of $0.1$ whereas the the solid curves are for the worst case where the instrument emissivity is $1.$ We also simulated a range of temperatures normally expected at the observatories. It is clear from the simulations that even in the best case, the thermal flux becomes the dominant noise component long ward of $J$-band. Therefore, the spectrograph can achieve sky-limited performance only short ward of 1.35 $\mu$m, provided there is a filter that blocks all light long ward of 1.35 $\mu$m. In the H-band, performance will be significantly degraded because the thermal light within the H-band bandpass cannot be blocked. However, our goal has been to optimize the spectrograph performance in the $0.9-1.35$ $\mu$m range and this issue is not of serious concern. 

\section{Predicted On-sky Performance}

Having finalized the optical design and the general mechanical characteristics of WIFIS, we simulated its on-sky performance in the $0.9-1.35$ $\mu$m range where the instrument sensitivity is not degraded by its own thermal light. Simulations were carried out for two different instrument destinations: (1) 2.3-meter Steward Bok telescope at Kitt Peak, USA, and (2) 10.4-meter Gran Telescopio Canarias (GTC) in the Canary Islands, Spain. These two possible destinations span the range of telescope aperture sizes that WIFIS can be used with and provide the upper and lower limits of WIFIS sensitivity. The system parameters for WIFIS derived from the telescope characteristics and the instrument's optical design are given in Table \ref{tab:parameters}. 
\begin{table}[h]
\caption{System parameters for WIFIS on two different telescopes.} 
\label{tab:parameters}
\begin{center}       
\small
\begin{tabular}{|l|l|l|} 
\hline
\rule[-1ex]{0pt}{3.5ex}  {\bf Telescope} & {\bf \emph{Steward Bok}}  & {\bf  \emph{GTC}} \\
\hline
\rule[-1ex]{0pt}{3.5ex}  {\bf Location} &  Kitt Peak, USA & La Palma, Spain \\
\hline
\rule[-1ex]{0pt}{3.5ex}  {\bf Diameter} & 2.3-meter & 10.4-meter   \\
\hline
\rule[-1ex]{0pt}{3.5ex}  {\bf Focal Ratio} & f/9 & f/17   \\
\hline
\rule[-1ex]{0pt}{3.5ex}  {\bf Typical Seeing} & $1.5-2^{\prime\prime}$ & $0.4-1^{\prime\prime}$   \\
\hline
\rule[-1ex]{0pt}{3.5ex}  {\bf Field-of-view} & $19.8^{\prime\prime}\times 50^{\prime\prime}$ &  $4.5^{\prime\prime}\times 12^{\prime\prime}$ \\
\hline
\rule[-1ex]{0pt}{3.5ex}  {\bf IFU Spatial Sampling} & $1.1^{\prime\prime}$/slice &  $0.25^{\prime\prime}$/slice \\
\hline
\rule[-1ex]{0pt}{3.5ex}  {\bf Detector Spatial Pixel Scale} & $0.55^{\prime\prime}\times0.66^{\prime\prime}$ &  $0.125^{\prime\prime}\times0.15^{\prime\prime}$ \\
\hline
\rule[-1ex]{0pt}{3.5ex}  {\bf Resolving Power} & $3,000$ & $3,000$  \\
\hline
\end{tabular}
\end{center}
\end{table} 

\par
\begin{figure}
   \begin{center}
   \begin{tabular}{c}
   \includegraphics[height=8cm]{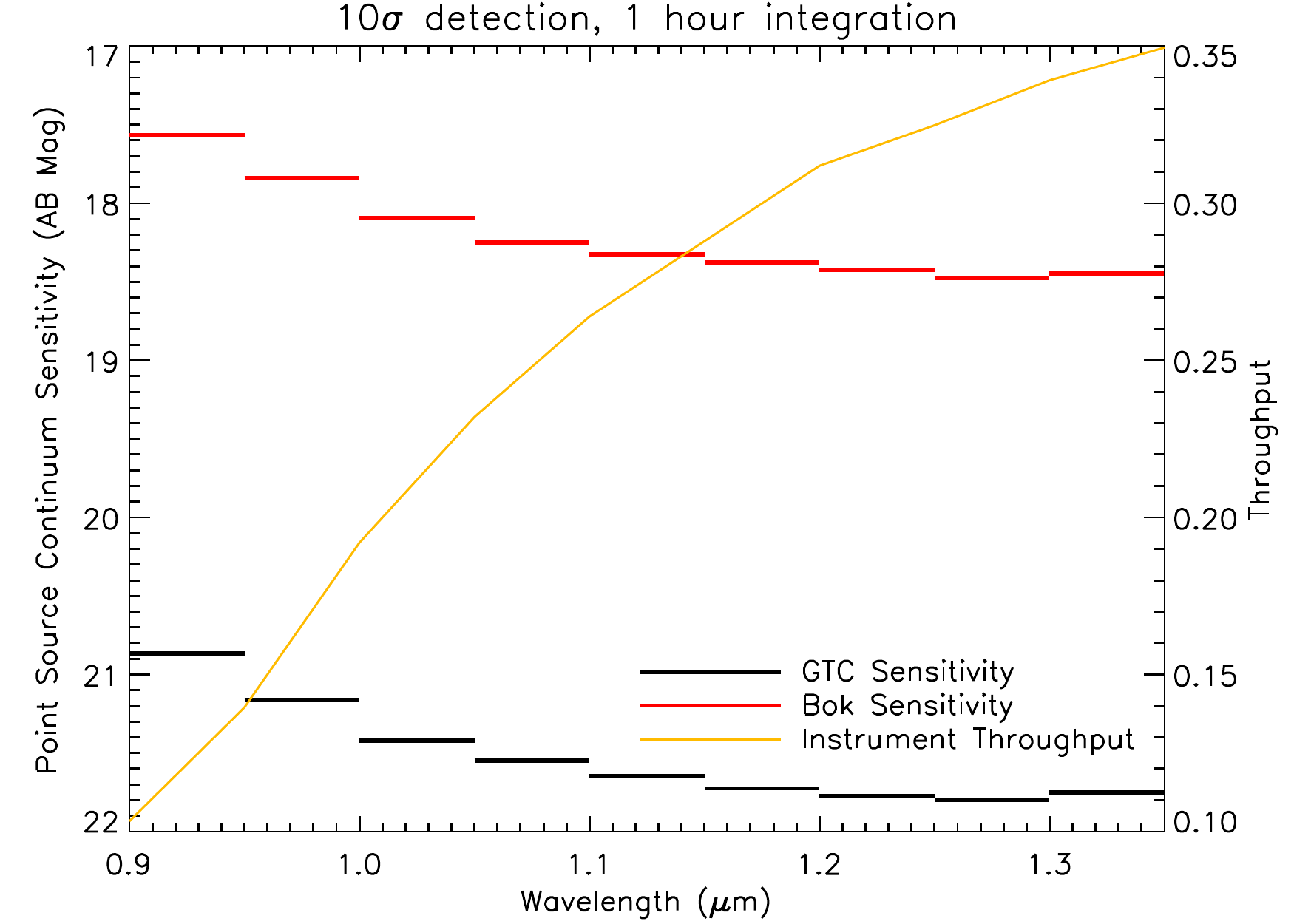} 
   \end{tabular}
   \end{center}
   \caption[contsensitivity]
   { \label{fig:contsensitivity} Predicted point source continuum sensitivity at the GTC (black curve) and the Bok (red curve) telescopes and instrument throughput (orange curve) as a function of wavelength. The plotted continuum sensitivity is for a 10$\sigma$ detection after an hour of integration time. The darkest regions in between the OH lines and typical seeing conditions are used to compute the sensitivity. The instrument throughput is also plotted, which does not include losses from the telescope.
   
}
   \end{figure}

End-to-end simulations that used realistic models of atmospheric transmission and background, throughput of the telescope and the instrument, and detector characteristics were used to predict the final sensitivity of the instrument. For modelling the atmospheric transmission and background, we obtained the necessary files from the Gemini Near-IR Sky Background\footnote[1]{http:$//$www.gemini.edu$/$?q=node$/$10787} and Transmission\footnote[2]{http:$//$www.gemini.edu$/$?q=node$/$10789} websites. We chose the files representative of Manua Kea during fair conditions: water column = 3.0 mm and air mass = 1.5. The sky brightness was rescaled to match a more realistic estimate of the sky brightness in between the OH lines in the $J$-band of 17.7 AB mag arcsec$^{-2}$\footnote[3]{http://irlab.astro.ucla.edu/mosfire/MOSFIRE\%20PDR\%20Report\%20v4.pdf}. To estimate the instrument throughput, we took into account the reflectivity of the mirrors, the transmission of the lenses and cryostat window, the blocking filter transmission, the average efficiency of the grating as a function of wavelength, and the detector quantum efficiency (QE). All mirrors including those within the IFU are gold-coated. We assume a reflectivity of 0.98 for these mirrors. The lenses and the window are AR-coated to reflect less than 1.5\% of their light and their internal transmission is close to 100\% across the wavelengths of interest. We assume a lens/window transmission of 0.985. For the thermal blocking filter, we assume a throughput of 0.8. The grating efficiency was taken to be the values provided by Newport for the corresponding master grating. The QE of the detector used for our simulations was given in its test report, and we present the QE values in Table \ref{tab:detector}. The overall throughput of the instrument is shown in Figure \ref{fig:contsensitivity}. Because the blaze wavelength of the grating is approximately $1.35$ $\mu$m, the throughput of the instrument is relatively low in the blue end, reaching 10\%. However, on the red end where we are closer to the blaze wavelength of the grating, we reach a much higher throughput of 35\%. Note that this does not include the telescope throughput, which needs to be included in the sensitivity calculations. For these calculations, we assumed the primary and secondary telescope mirrors were aluminum coated and each had a reflectivity of 0.92. 

\begin{table}[h]
\caption{HAWAII 2RG Detector Parameters.} 
\label{tab:detector}
\begin{center}       
\small
\begin{tabular}{|l|l|} 
\hline
\rule[-1ex]{0pt}{3.5ex}  {\bf Wavelength Range} & $0.8-2.5$ $\mu$m  \\
\hline
\rule[-1ex]{0pt}{3.5ex}  {\bf \# of Pixels} & $2048\times2048$  \\
\hline
\rule[-1ex]{0pt}{3.5ex}  {\bf Pixel size} &  18 $\mu$m  \\
\hline
\rule[-1ex]{0pt}{3.5ex}  {\bf Dark Current} & $< 0.02$ e$^{-}$ s$^{-1}$ pixel$^{-1}$   \\
\hline
\rule[-1ex]{0pt}{3.5ex}  {\bf Read Noise} & $5$ e$^{-}$   \\
\hline
\rule[-1ex]{0pt}{3.5ex}  {\bf Full Well Depth} & $120,000$ e$^{-}$   \\
\hline
\rule[-1ex]{0pt}{3.5ex}  {\bf Quantum Efficiency} & 0.75 @ 0.8 $\mu$m   \\
  & 0.76 @ 1.0 $\mu$m \\
  & 0.76 @ 1.2 $\mu$m \\
   & 0.84 @ 2.0 $\mu$m \\
\hline
\end{tabular}
\end{center}
\end{table} 
\par
\begin{figure}
   \begin{center}
   \begin{tabular}{c}
   \includegraphics[height=8cm]{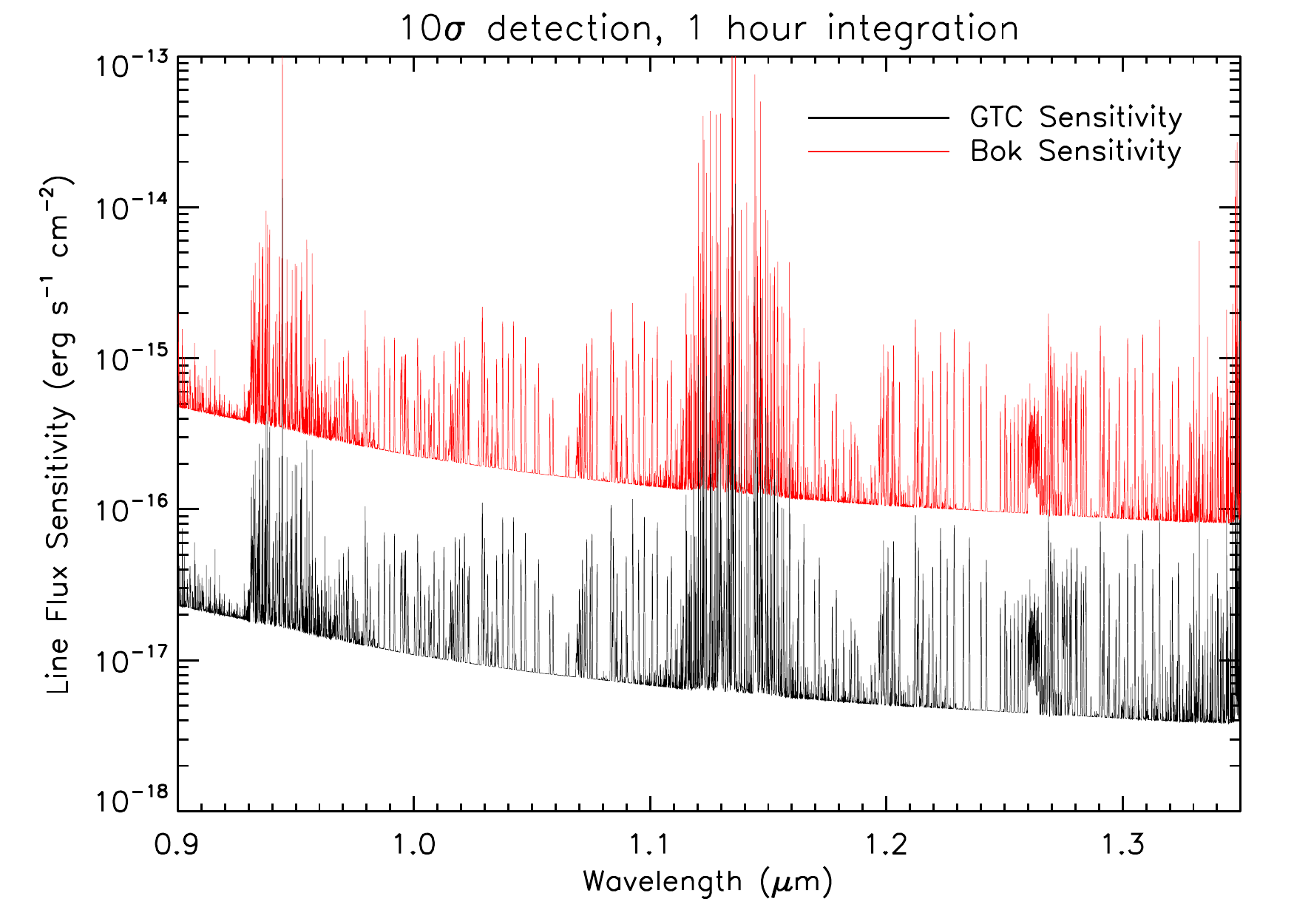} 
   \end{tabular}
   \end{center}
   \caption[linesensitivity]
   { \label{fig:linesensitivity} 
   Predicted unresolved line detection sensitivity at the GTC (black curve) and the Bok (red curve) as a function of wavelength. The calculation assumes the source is a point source with a width set to the typical seeing of each site. 
}
   \end{figure}

We estimated both the point source continuum and unresolved line sensitivity for the instrument at both the Bok and the GTC telescopes. The assumed detector noise characteristics are given in Table \ref{tab:detector}. The expected read noise can be achieved through 16-frame Fowler sampling. The assumed thermal photon flux was 0.5 s$^{-1}$ pixel$^{-1},$ which is the most conservative estimate when using a filter that cuts off all light long ward of 1.35 $\mu$m. The point source width was set to be 0.5\arcsecl at the GTC and 2\arcsecl at Bok, which are the typical seeing values at those sites. The predicted 10$\sigma$ continuum and unresolved line sensitivities for an hour long integration are shown in Figures \ref{fig:contsensitivity} and \ref{fig:linesensitivity}, respectively. The hour long integrations are taken as 10 six-minute exposures. We compare our sensitivities with a more complex seeing-limited, NIR, integral field spectrograph, KMOS, for the Very Large Telescope that will be commissioned later this year. The predicted 5$\sigma$ point source continuum sensitivity for the KMOS instrument is 24 AB mag in $1.0-1.5$ $\mu$m range for $R\sim 1,000$\cite{sharples04}. Rescaling to $R\sim3,000$ and for 10$\sigma$ detection, the KMOS sensitivity would then be 22 AB mag in $1.0-1.5$ $\mu$m range. At 1.0$\mu$m our GTC sensitivity is approximately 0.5 mag worse, but at 1.3$\mu$m our sensitivity is comparable. On the smaller aperture size end, the sensitivity of the TripleSpec NIR spectrograph at the 3.5-m Apache Point Observatory (APO) is a reasonable comparison. The measured 5$\sigma$ continuum sensitivity for this spectrograph is 18 AB mag in $J$-band. If we correct for the difference in aperture size between APO and Bok, we expect a 10$\sigma$ sensitivity of 16.3 AB mag for TripleSpec. This is almost $1-2$ mag worse than our predicted performance. This can be attributed to at least in part to the reduced throughput ($<0.1$) of TripleSpec in $J$-band\cite{herter08}. This difference in throughput could account for a 1.3 mag difference in sensitivity, but not 2 magnitudes. It may be possible that our expected performance at the Bok telescope is slightly optimistic, as the atmospheric transmission and background is likely worse at Kitt Peak when compared to Mauna Kea. 

\section{IFU Lab Characterization}

\begin{figure}
   \begin{center}
   \begin{tabular}{c}
   \includegraphics[height=8cm]{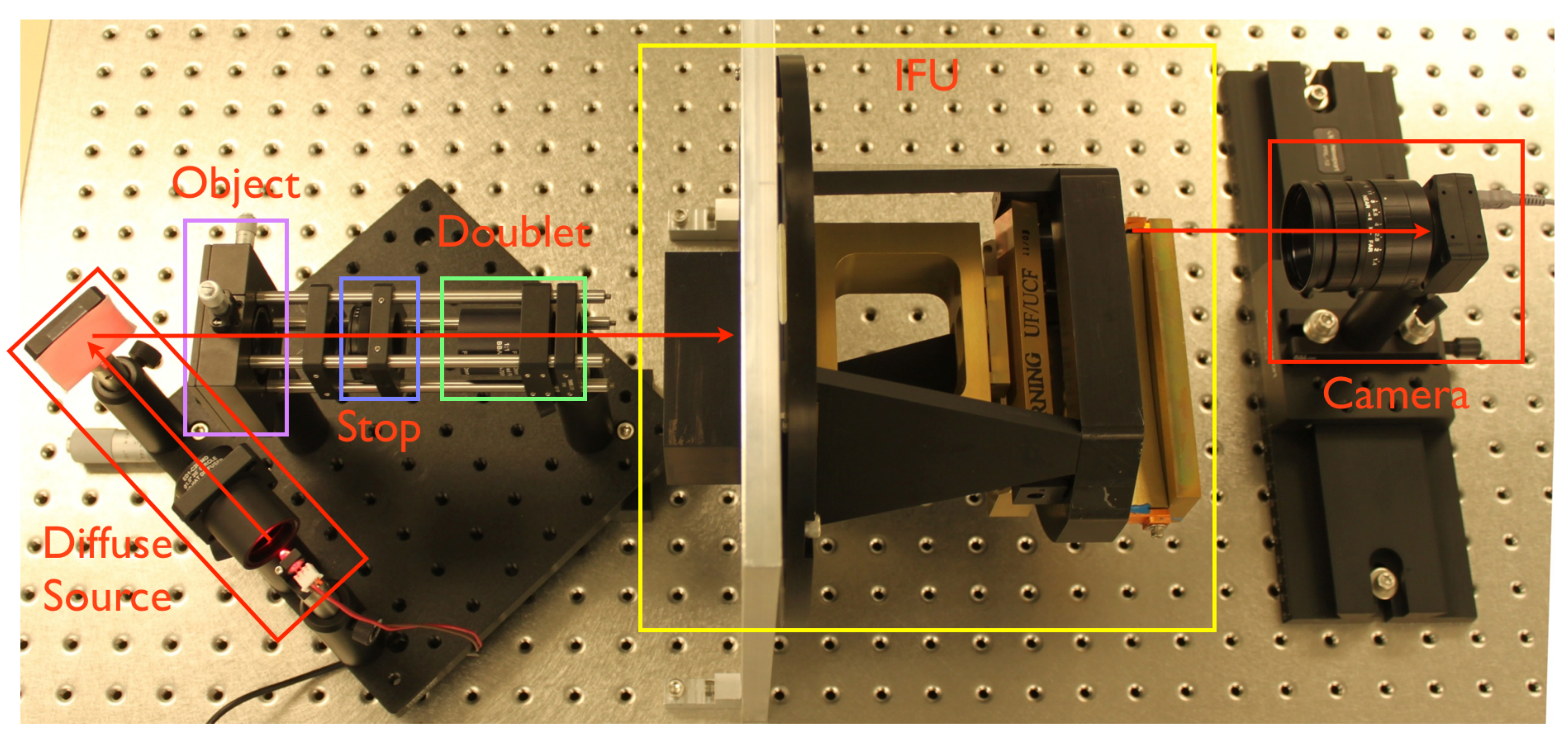} 
   \end{tabular}
   \end{center}
   \caption[labsetup]
   { \label{fig:labsetup} 
   Top view of the laboratory setup used to carry out the IFU tests. On the left on a separate optical breadboard is the telescope simulator which is mounted on a translation stage. The telescope simulator creates an extended source that overfills the IFU's field-of-view. Different objects such as pinholes and slits can be placed at the object translation stage. The stop size and position are set to produce an f/16 telecentric beam at the IFU's input. The virtual slit is imaged using a camera located behind (to the left) of the IFU. The light path is shown by the red arrows. The light path within the IFU is not shown for simplicity. 
      }
\end{figure}

To characterize the optical properties of FISICA, we constructed a laboratory test setup that simulated a telescope beam over a sufficiently large field to fill the IFU input field. The main IFU properties that were measured by our tests were its magnification, field mapping, and its virtual slit characteristics. Specifically, it was necessary to measure the slit width along the length of the virtual slit in order to confirm the instrument will achieve the necessary spectral resolution across its full field-of-view. In order to accomplish this, we constructed a telescope simulator that produced an f/16 telecentric beam with an image plane that was coincident with the IFU's input field. The test setup used for this purpose is shown in Figure \ref{fig:labsetup}. The simulator was an optical system that consisted of a diffuse source, an object (in the form of a pinhole or slit), an adjustable stop, and an off-the-shelf achromatic doublet. The diffuse source consisted of a bright 650 nm light emitting diode (LED) source, an engineered diffuser, and a lambertian diffuser. The engineered diffuser, located immediately after the LED source, produced relatively uniform illumination across a 20$^\circ$ solid angle. The light emanating from the engineered diffuser was scattered by a lambertian diffuser, which illuminated the object. The object was imaged onto the IFU field by the achromatic doublet. The location and size of the adjustable stop was set to produce an f/16 telecentric output beam. The entire telescope simulator was on a translation stage that allows one to adjust the image plane of the simulator to be coincident with the IFU input field. The virtual slit was then imaged by a camera located behind the IFU. This camera was translated along the slit direction to capture the full slit. 

\begin{figure}
   \begin{center}
   \begin{tabular}{c}
   \includegraphics[height=9cm]{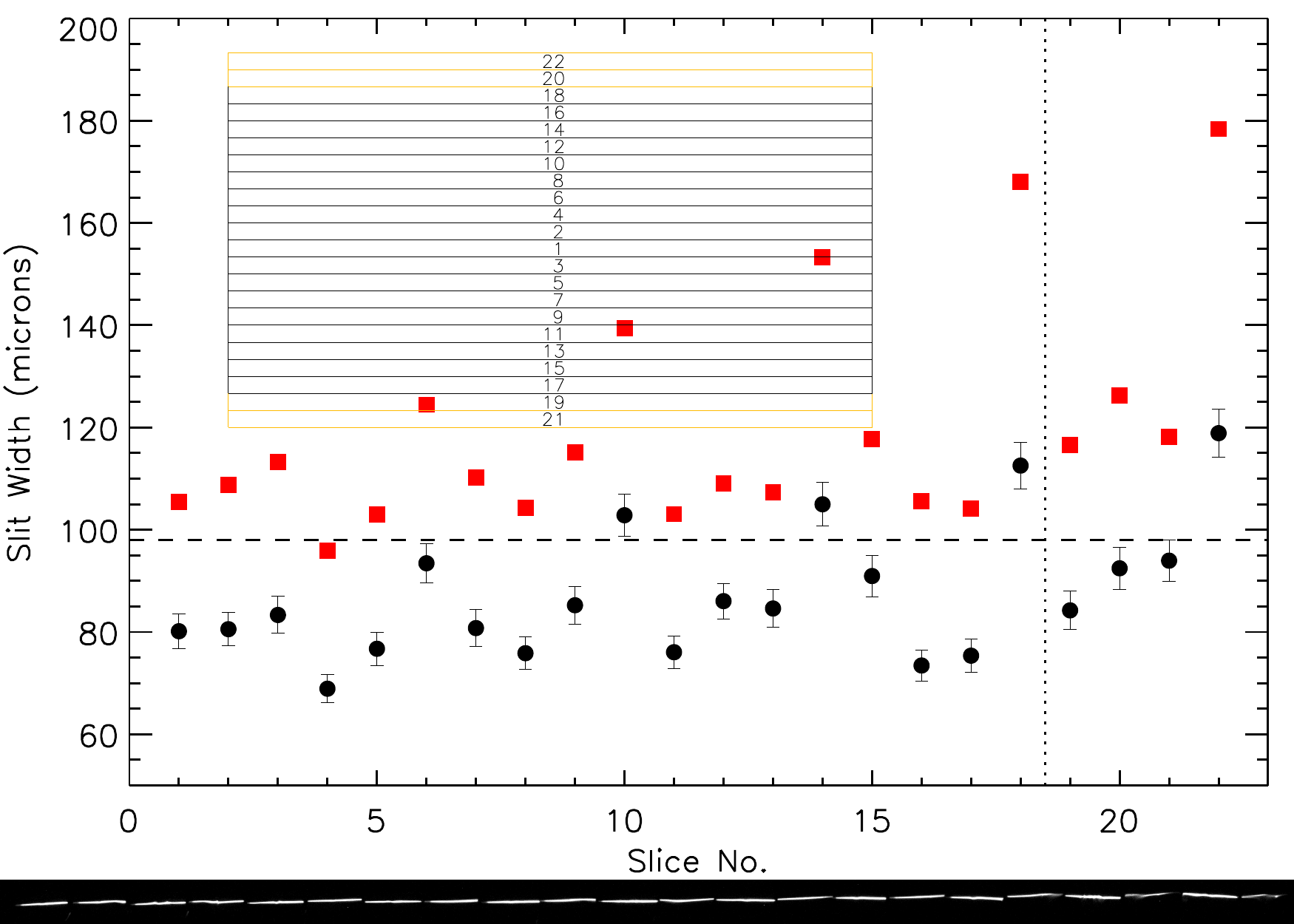} 
   \end{tabular}
   \end{center}
   \caption[slitwidth]
   { \label{fig:slitwidth} 
   {\it Top:} Average slit width as a function of slice number. Slit width is measured using two different techniques. The first technique uses the FWHM derived from fitting a Gaussian function to the line spread function (black points). The second method measures the width within which 80\% of the power is enclosed (red points). The inset in the upper left corner of this plot shows how the telescope field is mapped to slice number. The top two slices and the bottom two slices (orange) are not used in our spectrograph. The dashed horizontal line is the nominal slit width we expect for the ideal operation of the IFU and is also the value used for our spectral resolving power simulations. The slices that are located to the right of the dotted vertical line are not used in our spectrograph. {\it Bottom:} A virtual slit image taken by our test setup. Each roughly horizontal line represents each slice, with slice number \#1 located at the very left and slice number \#22 located on the very right.
   }
\end{figure}

The various IFU parameters such as its magnification and field-mapping were confirmed by our testing. To understand the intrinsic slit width of the IFU, we measured the virtual slit width as a function of slit position. This is an important measurement that will partly determine the final spectral resolution of our spectrograph. Unlike a typical long slit spectrograph that has an entrance aperture that resembles a top hat function with a width corresponding to a slit width, the slit image produced by the image slicer has slightly different properties. The slit line spread function has a core that can be modelled well by a Gaussian, but has a more extended tail possibly due to light scattered off the edges of the slicer. We present the results of our slit width measurement along with an image of a virtual slit in Figure \ref{fig:slitwidth}. The slit widths were determined using two different methods. The first method fit a Gaussian function to the slit, and the width of the slit was measured to be the FWHM of the Gaussian. In general, the measured width is less than the expected slit width of 98 $\mu$m. While the slit line spread function was reasonably well fit by a Gaussian function at its core, the wings of the line spread function were fit poorly by a Gaussian because there was more power in the wings. This was most likely from scattered light from the IFU, as one might expect the image slicer does not have perfectly hard transitions between different slices. Another metric to measure the slit width is to determine the width within which 80\% of the light from the line spread function is contained, hereafter called EE80. The average value of this metric for each slice is also shown in Figure \ref{fig:slitwidth}. For the 18 slices that are used in the WIFIS design, the average value of EE80 is 116 $\mu$m, which is slightly worse than the expected slit width. It is worth noting, however, that there are a few slices with much worse performance, ranging from $140-170$ $\mu$m.

\section{Current Status}
The optical design of the instrument has been completed and we are in the process of acquiring all of the optical components and the detector. There are a number of remaining tasks: (1) the design and fabrication the thermal blocking filter; (2) the design and fabrication of the mechanical components such as the optical bench, enclosure, lens and mirrors mount, and the cryostat; and (3) the assembly and testing of the instrument in the lab. We are currently characterizing optical components such as the reflective components, the gratings, and the IFU. Our plan is to complete the assembly and testing of the instrument by early 2013 and bring the instrument to the Bok telescope during Spring 2013 for commissioning followed by regular scientific observations starting in Summer 2013.

\section{Conclusions}
In this work, we present the final optical design and system layout of WIFIS, a wide integral field IR spectrograph, designed to operate in the $0.9-1.8$ $\mu$m wavelength range. WIFIS has an unrivalled etendue when it comes to observing a single field in the NIR. This is achieved through the use of image-slicer type IFU, complex optics, and a large format H2RG detector array. The wide-field capability of WIFIS benefits a unique class of scientific problems, previously inaccessible by small-field NIR integral field spectrographs, that range from ionization and chemical properties of nearby star forming regions to kinematics and star forming properties of moderate redshift galaxies. The versatility of WIFIS also allows one to use the instrument in different classes of telescopes ranging from a 2.3-meter telescope to a 10-meter one. Each telescope destination provides a unique set of capabilities because the field-of-view of the instrument scales inversely with telescope size even though the sensitivity is reduced at a smaller telescope. We also carry out simulations to determine the predicted on-sky sensitivity and spectral resolving power of the instrument at the Bok and GTC. We find that our sensitivities are comparable to other similar instruments that will be commissioned in the near future. Finally, we present results from our laboratorial characterization of the integral field unit. We plan on commissioning this instrument at the Steward Bok telescope during Spring 2013.

\acknowledgments     
 
S.S. was supported by the Dunlap Institute through the Dunlap Fellowship program. This project has been largely supported by Leaders Opportunity Fund (13038) of the Canadian Foundation for Innovation and by Ontario Research Fund of the Ministry of Research and Innovation of the Ontario Provincial Government to D.-S. Moon.

\bibliography{report}   
\bibliographystyle{spiebib}   

\end{document}